\documentclass[review]{elsarticle}

\usepackage{hyperref}

\journal{Physics Letters B}

\def\bear{\begin{eqnarray}}
\def\ear{\end{eqnarray}\noindent}

\renewcommand{\theequation}{\arabic{equation}}
\newcommand{\be}{\begin{equation}}
\newcommand{\ee}{\end{equation}}
\newcommand{\ea}{\end{array}}
\newcommand{\beqa}{\begin{eqnarray}}
\newcommand{\eeqa}{\end{eqnarray}}

\newcommand{\e}{\mathrm{e}}

\def\blue{\color{blue}}
\def\bear{\begin{eqnarray}}
\def\ear{\end{eqnarray}\noindent}
\def\bec{\blue\begin{equation}}
\def\eec{\end{equation}\black\noindent}
\def\bearc{\blue\begin{eqnarray}}
\def\earc{\end{eqnarray}\black\noindent}
\def\benn{\begin{enumerate}}
\def\enn{\end{enumerate}}

\def\half{\frac{1}{ 2}}

\def\slash#1{#1\!\!\!\raise.15ex\hbox {/}}

\def\t{\tau}
\def\veps{\varepsilon}

\def\4piTD{{(4\pi T)}^{-{D\over 2}}}
\def\4piT4{{(4\pi T)}^{-2}}

\def\Tintm4{{\dps\int_{0}^{\infty}}{dT\over T}\,e^{-m^2T}
    {(4\pi T)}^{-2}}
\def\Tintm{{\dps\int_{0}^{\infty}}{dT\over T}\,e^{-m^2T}}

\newcommand{\slG}{{{\dot G}\!\!\!\! \raise.15ex\hbox {/}}}

\def\GBd12{{\dot G}_{B12}}

\newcommand{\no}{\noindent}
\def\non{\nonumber \\}
\def\dps{\displaystyle}

\def\O(#1){O($T^#1$)} 
\def\O2{O($T^2$)}
\def\O3{O($T^3$)}
\def\O4{O($T^4)}
\def\O5{O($T^5$)}

\def\no{\noindent}

\begin{document}

\begin{frontmatter}

\title{Off-shell Ward identities for N-gluon amplitudes}

\author{Naser Ahmadiniaz\fnref{mycorrespondingauthor}}
\ead{n.ahmadiniaz@hzdr.de}
\address{Helmholtz-Zentrum Dresden-Rossendorf, Bautzner Landstra\ss e 400, 01328 Dresden, Germany}
\fntext[mycorrespondingauthor]{Corresponding author}


\author{Christian Schubert}
\address{Instituto de F{{\'\i}}sica y Matem\'aticas, Universidad Michoacana de San Nicol\'as de Hidalgo\\
Apdo. Postal 2-82, C.P. 58040, Morelia, Michoacan, Mexico}
\ead{schubert@ifm.umich.mx}

\begin{abstract}
Off-shell Ward identities in non-abelian gauge theory continue to be a subject of active research,
since they are, in general, inhomogeneous and their form depends on the chosen gauge-fixing procedure. 
For the three-gluon and four-gluon vertices, 
it is known that a relatively simple form of the Ward identity 
can be achieved using the pinch technique or, equivalently, the background-field method with quantum Feynman gauge. The latter is also the gauge-fixing underlying the string-inspired formalism, and here we use this formalism to derive the corresponding form of the Ward identity
for the one-loop N - gluon amplitudes. 
\end{abstract}


\end{frontmatter}


\section{Introduction: Ward-Takahashi and Slavnov-Taylor identities}
\label{sec:intro}
\renewcommand{\theequation}{1.\arabic{equation}}
\setcounter{equation}{0}

Ward identities, also known as Ward-Takahashi identities, are the quantum counterparts to Noether's theorem in classical physics. They are identities between correlation functions stemming
from the global and gauge symmetries of the theory, introduced first by Ward \cite{ward} in 1950 and later generalized by Takahashi \cite{takahashi}. 

The original work of Ward and Takahashi was concerned with 
$U(1)$ gauge symmetry and current conservation in QED. Here the ``Ward identity'' refers to on-shell matrix elements, and is usually
written as 
\bear
k^\mu\mathcal{M}_\mu=0\,,
\label{ward}
\ear
where $\mathcal{M}_\mu$ is the matrix element defined by $\mathcal{M}=\varepsilon_\mu\mathcal{M}^\mu$, indicating
that the longitudinal components of the photon's polarizations  do not contribute to scattering amplitudes. 
See, e.g., \cite{peskin} for detailed discussion. 

The ``Ward-Takahashi'' identity is more involved, since it concerns off-shell quantities. In QED, in its most basic form it 
can be written as
\bear
\Gamma_{\mu}(p,p) = - \frac{\partial}{\partial p^{\mu}} \Sigma (p)\,,
\label{wt}
\ear
and allows one to relate the electron wave function renormalization factor $Z_2$ to the vertex renormalization factor $Z_1$.

After the development of QCD, in the seventies the generalization of these QED identities to the non-abelian case became an active field of research.
With respect to the on-shell S-matrix identities (\ref{ward}) one finds no essential differences between QED and QCD, except that in the non-abelian case the vanishing
of the effect of the longitudinal gluon polarizations usually involves intricate cancellations between one-particle irreducible and one-particle reducible diagrams (see, e.g., \cite{leapre}). 

To the contrary, the generalization of the Ward-Takahashi identities to the non-abelian case leads to the so-called
Slavnov-Taylor identities \cite{taylor-71,slavnov-72,pastar-book}, and those still remain a subject of active investigation (see, e.g., \cite{grkikr} and refs. therein). 
This is because these identities in general not only provide relations between different $N$-point functions,
but also mix up the physical gauge bosons with the ghosts, and in a way that depends on the gauge-fixing procedure. 
Thus they tend to be much more non-trivial than their QED prototype. 
Moreover, they should hold perturbatively and non-perturbatively, and in the bare theory as well as in the renormalized one. Therefore Slavnov-Taylor identities also put restrictions on the renormalized coupling constants for vertex functions which has been studied in detail by many authors, see e.g. \cite{thooft-71,ball-chiu-80,kim-baker-80}. 

The gauge-fixing dependence constitutes a serious problem for applications of the Schwinger-Dyson equations in QCD.
Those equations couple an infinite number of Green's functions, and attempts at explicit solution normally require a
truncation to a finite subset of them. This truncation should be gauge-invariant, which in the non-abelian case is not
easily achieved. This triggered the development of the ``pinch technique'' (`PT') \cite{cornwall,binpap-rev}, a systematic procedure that allows
one to construct, starting from the standard Green's functions derived from the gauge-fixed QCD Lagrangian,
improved ``gauge-invariant'' vertices that fulfill simple QED-like off-shell Ward identities, not involving the ghosts. 
For the $N$-gluon vertices, which are our subject of interest in this letter, this procedure has, to the best of our knowledge,
been carried out only for $N=3$ and $N=4$, and only at the one-loop level. The three-point vertex is special in that it
involves the color indices only as a global prefactor:

\bear
\Gamma_{\mu_1\mu_2\mu_3}^{abc} (k_1,k_2,k_3)= -if^{abc} \Gamma_{\mu_1\mu_2\mu_3} (k_1,k_2,k_3)\,,
\ear
where the $f_{abc}$ are the structure constants of the Lie algebra, $[T^a,T^b]=iT^cf^{abc}$. 
As shown by Cornwall and Papavassiliou \cite{corpap}, when constructed
using the PT it will obey the identity 

\bear
k_1^{\mu_1} \Gamma_{\mu_1\mu_2\mu_3} (k_1,k_2,k_3)
&=& 
-(k_2^2 g_{\mu_2\mu_3} - k_{2\mu_2}k_{2\mu_3}) \Bigl(1-\Pi(k_2^2)\Bigr)
\nonumber\\
&&
+(k_3^2 g_{\mu_2\mu_3} - k_{3\mu_2}k_{3\mu_3}) \Bigl(1-\Pi(k_3^2)\Bigr)\,,
\nonumber\\
\label{ward3gluon}
\ear
where $\Pi (k^2)$ is the gluon vacuum polarization. 
This form of the Ward identity holds unambiguously for the scalar and spinor loop cases, but for the
gluon loop with other gauge fixings in general there will be additional terms on the right-hand-side involving 
not only the gluon propagator, but also the ghost propagator and the gluon-ghost-ghost vertex \cite{ball-chiu-80,marpag,binbro}. The corresponding identity for the four-gluon vertex
was given by Papavassiliou  \cite{papavassiliou4gluon}

\bear
k_1^{\mu_1}\Gamma_{\mu_1\ldots \mu_4}^{a_1a_2a_3a_4}(k_1,k_2,k_3,k_4)
&=&-ig f_{a_1a_2c}\Gamma^{ca_3a_4}_{\mu_2\mu_3\mu_4}(k_2+k_1,k_3,k_4)
\nonumber\\
&& 
-ig f_{a_1a_3c}\Gamma^{a_2ca_4}_{\mu_2\mu_3\mu_4}(k_2,k_3+k_1,k_4)
\nonumber\\
&&
-ig f_{a_1a_4c}\Gamma^{a_2a_3c}_{\mu_2\mu_3 \mu_4}(k_2,k_3,k_4+k_1)\,.
\nonumber\\
\label{ward4gluon}
\ear
From a somewhat different perspective, the issue of the gauge-fixing dependence of the off-shell Ward identities  
arises also in the context of SUSY extensions of QCD \cite{bddk94}, and was studied in detail for the
three-gluon vertex by Binger and Brodsky \cite{binbro}. 
Here as a minimal requirement one would wish to have a gauge-fixing procedure compatible with the supersymmetry, 
in particular with the structure of supermultiplets; e.g., one would like to avoid having to use different Ward
identities for amplitudes that differ, say, only by different components of the same superfield running in a loop. 
For this purpose, the background-field method (`BFM') 
\cite{abbott,abgrsc} turned out to be very suitable, and in fact equivalent to
the PT, since it was shown in \cite{dewedi,papavassiliou95} that the application of the BFM with quantum Feynman gauge leads to exactly the same Green's functions as the PT.

The BFM with quantum Feynman gauge is also the formalism underlying the construction of the 
one-loop $N$-gluon amplitudes in 
the ``string-inspired worldline formalism'' \cite{strassler1,strassler2}
which to some extent mimics the construction of
gauge boson amplitudes in string perturbation theory.
This formalism therefore is guaranteed to lead to the
same simple and ghost-free, QED-like off-shell Ward identities as the PT. 
Moreover, one of the properties that it inherits from string theory is that it allows one to unify
the scalar, spinor and gluon-loop contributions to these amplitudes in a way that would be difficult to achieve in other
approaches, namely by a set of simple pattern-matching rules at the parameter-integral level 
due to Bern and Kosower \cite{strassler1,berkos-npb362,berkos-npb379}.
These rules were originally derived from world-sheet SUSY, but can also be derived by more direct means \cite{18,41}.
In the present letter, we will use these properties to show that
the identities (\ref{ward3gluon}), (\ref{ward4gluon}) generalize to the $N$-gluon case in the simplest possible way,
namely as

\bear
k_1^{\mu_1}\Gamma_{\mu_1\ldots \mu_N}^{a_1a_2\cdots a_N}(k_1,\ldots,k_N)
&=&-ig f_{a_1a_2c}\Gamma^{ca_3a_4\cdots a_N}_{\mu_2\ldots \mu_N}(k_2+k_1,k_3,\cdots,k_N)
\nonumber\\
&& 
-ig f_{a_1a_3c}\Gamma^{a_2ca_4\cdots a_N}_{\mu_2\ldots \mu_N}(k_2,k_3+k_1,\cdots,k_N)
\nonumber\\
&& \qquad  \vdots \nonumber\\
&&
-ig f_{a_1a_Nc}\Gamma^{a_2a_3a_4\cdots c}_{\mu_2\ldots \mu_N}(k_2,k_3,\cdots,k_N+k_1)\,.
\nonumber\\
\label{wardNgluon}
\ear

\section{String-inspired representation of gluon amplitudes}
\label{sec:si}
\renewcommand{\theequation}{2.\arabic{equation}}
\setcounter{equation}{0}

Around 1990, Bern and Kosower used the field theory limit of string theory to derive new
parameter integral representations for the QCD one-loop $N$-gluon amplitudes
\cite{berkos-npb362,berkos-npb379}.
In its original form this formalism was restricted to on-shell matrix elements, but 
it was soon extended to the off-shell case by Strassler using worldline path integral representations
of the same amplitudes \cite{strassler1, strassler2,18,5}. More recently,  this version of the formalism 
has been found particularly suitable for the study of non-abelian form factor decompositions \cite{91,92,105}.

Let us briefly summarize how the one-loop off-shell 1PI $N$-gluon amplitudes are constructed in the
string-inspired formalism for a scalar, spinor or gluon loop (for details see \cite{18}).
The starting point is the following path-integral representation of the scalar loop contribution to this amplitude:
\bear
\Gamma_{\rm scal}(k_1,\veps_1,a_1;\cdots; k_N,\veps_N,a_N)
&=&(ig)^N\int_0^\infty \frac{dT}{T}
\, e^{-m^2T}\int \mathcal{D}x\, e^{-\int_0^Td\tau \frac{1}{4}\dot{x}^2}
\nonumber\\
&& \times 
\, V_{\rm scal}[k_1,\veps_1,a_1]\cdots V_{\rm scal}[k_N,\veps_N,a_N]\,.\non
\label{wi-N}
\ear
Here $m$ is the mass and $T$ the proper-time of the scalar in the loop.
At fixed $T$, each gluon is represented by a vertex operator
\bear
V_{\rm scal}[k_i,\varepsilon_i,a_i] = T^{a_i}\int_0^T d\tau \varepsilon_i\cdot \dot x_i \,\e^{i k_i\cdot x_i}\,,
\label{defV}
\ear
where $k_i$ and $\varepsilon_i$ are the gluon momentum and polarization, $T^{a_i}$ denotes
a generator of the color group, and we have abbreviated
$\dot x_i \equiv \frac{d}{d\tau} x(\tau_i)$.
The polarization vectors $\varepsilon_i$ at this stage are just book-keeping devices, and do not fulfill any 
on-shell constraints. 


This is the full amplitude; 
each gluon vertex operator is integrated along the loop independently,
so that the color generators $T^{a_i}$ appear under the color trace in all possible orderings. 
It will be sufficient to consider the contribution corresponding to the standard ordering
$\tau_1 \geq \tau_2 \geq \ldots \geq \tau_N$, to be denoted by $\Gamma_{\rm scal}^{a_1\ldots a_N}$. 
It can be written as
\bear
\Gamma_{\rm scal}^{a_1\ldots a_N}
(k_1,\veps_1;\cdots; k_N,\veps_N)
&=&(ig)^N\int_0^\infty \frac{dT}{T}
\, e^{-m^2T}\int \mathcal{D}x\, e^{-\int_0^Td\tau \frac{1}{4}\dot{x}^2}
\nonumber\\
&& \qquad \times 
\, V_{\rm scal}[k_1,\veps_1,a_1]\cdots V_{\rm scal}[k_N,\veps_N,a_N]
\nonumber\\
&& \qquad \times 
\theta(\tau_1-\tau_2)
\theta(\tau_2-\tau_3)
\cdots
\theta(\tau_{N-1}-\tau_{N})
\delta(\frac{\tau_N}{T})\,.
\nonumber\\
\label{Ngluonscalarordered}
\ear
Apart from imposing the proper-time ordering, one can use here also the translation invariance in proper-time
to reduce the number of integrations by setting $\tau_N=0$.  The full amplitude (\ref{wi-N})
is obtained from the ordered one (\ref{Ngluonscalarordered}) by summing over all $(N-1)!$ inequivalent
permutations. 

In the string-inspired formalism the path integral (\ref{Ngluonscalarordered}) is done by gaussian
integration, leading to the following ``Bern-Kosower master formula'':
\bear
\Gamma_{\rm scal}^{a_1\ldots a_N}
(k_1,\veps_1;\cdots; k_N,\veps_N)
&=&
{(ig)}^N
{\rm tr}\bigl(T^{a_1}T^{a_2} \cdots T^{a_N}\bigr)(2\pi)^Di\delta^D\Big(\sum k_i\Big)
\nonumber\\
&& 
\hskip -3cm
\times {\dps\int_{0}^{\infty}}
dT
{(4\pi T)}^{-{D\over 2}}
e^{-m^2T}
\int_0^Td\tau_1\int_0^{\tau_1}d\tau_2 \ldots \int_{\tau_N=0}^{\tau_{N-2}}d\tau_{N-1}
\nonumber\\
&& 
\hskip -3cm
\times
\exp\biggl\lbrace\sum_{i,j=1}^N 
\Bigl\lbrack  \half G_{Bij} k_i\cdot k_j
-i\dot G_{Bij}\varepsilon_i\cdot k_j
+\half\ddot G_{Bij}\varepsilon_i\cdot\varepsilon_j
\Bigr\rbrack\biggr\rbrace
\mid_{\rm multi-linear} \,.
\nonumber\\
\label{BKmaster}
\ear
Here $D$ is the space-time dimension, and the notation 
$\mid_{\rm multi-linear}$ 
means that, after the expansion of the exponential, only terms 
linear in every polarization vector should be retained. 
$G_{Bij}$ stands for the ``bosonic worldline Green's function''
\be
G_B(\tau_i,\tau_j)=\mid \tau_i-\tau_j\mid 
-{{(\tau_i-\tau_j)}^2\over T} - \frac{T}{6}\,.
\label{defG}
\ee
Writing out the exponential in eq.(\ref{BKmaster})
one obtains an integrand
\be \exp\biggl\lbrace 
\cdot
\biggr\rbrace \mid_{\rm multi-linear} 
\quad={(-i)}^N P_N(\dot G_{Bij},\ddot G_{Bij})
 \exp\biggl[\half \sum_{i,j=1}^N G_{Bij}k_i\cdot
k_j \biggr]\,, \label{defPN} \ee\noindent 
with a certain polynomial $P_N$.

Starting from this parameter-integral representation of the scalar loop contribution 
to the $N$-gluon amplitude, one can now generate the contributions
of the spinor and gluon loop in the following way. There exists a systematic
integration-by-parts procedure that eliminates all second derivatives of $G_B$ 
\cite{berkos-npb362,berkos-npb379,91,26}. 
After this, a parameter-integral representation of the spinor loop contribution can 
(up to the global normalization)  be generated from the scalar-loop one by replacing 
every ``$\tau$-cycle'' appearing in $Q_N$,
that is, a product of $\dot G_B$ whose indices form a cycle, 
according to the ``cycle-replacement rule''
\begin{eqnarray}
\dot G_{Bi_1i_2} 
\dot G_{Bi_2i_3} 
\cdots
\dot G_{Bi_ni_1}
\rightarrow 
\dot G_{Bi_1i_2} 
\dot G_{Bi_2i_3} 
\cdots
\dot G_{Bi_ni_1}
-
G_{Fi_1i_2}
G_{Fi_2i_3}
\cdots
G_{Fi_ni_1}\,,
\nonumber\\
\label{subrule}
\end{eqnarray}
\no
where $G_{F12}\equiv {\rm sign}(\tau_1-\tau_2)$
denotes the `fermionic' worldline Green's
function. 
A similar ``cycle replacement rule'' allows one to generate a parameter-integral
representation of the gluon-loop contribution.

For our present purpose it will further be important that in the partially integrated integrand
each $\tau$-cycle 
$\dot G_{Bi_1i_2} \dot G_{Bi_2i_3} \cdots \dot G_{Bi_ni_1}$
appears multiplied with a corresponding ``Lorentz-cycle''
${\rm tr}(f_{i_1}f_{i_2} \cdots f_{i_n})$, where 
\bear
f_i^{\mu\nu}&\equiv&
k_i^{\mu}\varepsilon_i^{\nu}
- \varepsilon_i^{\mu}k_i^{\nu}\,,
\label{deffi}
\ear
is the field-strength tensor of gluon $i$.

\section{Derivation of the N-gluon Ward identity: scalar loop}
\label{sec:derivation}
\renewcommand{\theequation}{3.\arabic{equation}}
\setcounter{equation}{0}

Let us now turn to the Ward identity in the $N$-gluon case. 
Starting from the path-integral representation (\ref{wi-N}) of the 
scalar contribution to the $N$-gluon amplitude, 
and replacing $\veps_i$ by $k_i$, the corresponding vertex operator becomes the integral of a total derivative,
and collapses to boundary terms:
\bear
V_{\rm scal}[k_i,\veps_i]&\stackrel{\veps_i\rightarrow k_i}{=}&T^{a_i}\int_{\tau_{i+1}}^{\tau_{i-1}}d\tau_i\,k_i\cdot \dot{x}(\tau_i)\,e^{ik_i\cdot x(\tau_i)}
\nonumber\\
&=&-iT^{a_i}\int_{\tau_{i+1}}^{\tau_{i-1}}d\tau_i \frac{\partial}{\partial \tau_i}\,e^{ik_i\cdot x(\tau_i)}\non
&=& -iT^{a_i}\Big[e^{ik_i\cdot x(\t_{i-1})}-e^{ik_i\cdot x(\t_{i+1})}\Big]\,.\non
\ear
After plugging this back it into eq. (\ref{wi-N}) for the natural ordering $\t_1\ge\t_2\ge\cdots\t_{i-1}\ge\t_i\ge\t_{i+1}\cdots\ge\t_N$
we have (in the following we omit the global energy-momentum conservation factor)

\bear
&&\Gamma^{a_1\cdots a_N}_{\rm scal}[k_1,\veps_1;\cdots; k_N,\veps_N]\stackrel{\veps_i\rightarrow k_i}{=}-i(ig)^N\, {\rm tr}(T^{a_1}\cdots T^{a_{i-1}}T^{a_i}T^{a_{i+1}}\cdots T^{a_N})\, \int_0^\infty \frac{dT}{T}\, e^{-m^2T}\non
&&\times\int \mathcal{D}x\, e^{-\int_0^Td\tau \frac{1}{4}\dot{x}^2}\bigg\{\int_0^Td\t_1\veps_1\cdot \dot{x}(\t_1)e^{ik_1\cdot x(\t_1)}\cdots \int_{0}^{\t_{i-2}}d\t_{i-1}\veps_{i-1}\cdot \dot{x}(\t_{i-1})e^{i(k_{i-1}+k_i)\cdot x(\t_{i-1})}\non
&&\times \int_{0}^{\t_{i-1}}d\t_{i+1}\veps_{i+1}\cdot \dot{x}(\t_{i+1})\, e^{ik_{i+1}\cdot x(\t_{i+1})}\cdots \int_{0}^{\t_{N-1}}d\t_N\veps_N\cdot \dot{x}(\t_N)e^{ik_N\cdot x(\t_N)}\non
&&\hspace{3cm}-\int_0^Td\t_1\veps_1\cdot \dot{x}(\t_1)e^{ik_1\cdot x(\t_1)}\cdots \int_{0}^{\t_{i-2}}d\t_{i-1}\veps_{i-1}\cdot \dot{x}(\t_{i-1})e^{ik_{i-1}\cdot x(\t_{i-1})}\non
&&\times \int_{0}^{\t_{i-1}}d\t_{i+1}\veps_{i+1}\cdot \dot{x}(\t_{i+1})\, e^{i(k_i+k_{i+1})\cdot x(\t_{i+1})}\cdots \int_{0}^{\t_{N-1}}d\t_N\veps_N\cdot \dot{x}(\t_N)e^{ik_N\cdot x(\t_N)}\bigg\}\,.\non
\label{wi-N1}
\ear
Let us now focus on the term from the lower boundary $\tau_i = \tau_{i+1}$. If we 
apply the same replacement to the ordering that differs from the standard one only by an 
exchange of $\tau_i$ and $\tau_{i+1}$, 
$\t_1\ge\t_2\ge\cdots\t_{i-1}\ge\t_{i+1}\ge\t_{i}\ge\cdots\t_N$, the same term will be generated from the
upper boundary of the $\tau_i$ integral, only with the opposite sign and an interchange of the 
color matrices $T^{a_i}$ and $T^{a_{i+1}}$. 
Thus in the abelian case all the boundary terms would cancel in pairs, 
but in the non-abelian theory instead each pair produces a color commutator. 
Inserting the $i$th vertex operator in all $N$ possible ways, but keeping the order of the
other vertex operators fixed, it is then easy to arrive at (\ref{wardNgluon}) (where we
have now set $i=1$). 

\section{Spinor and gluon loop}
\label{sec:spinorgluon}
\renewcommand{\theequation}{4.\arabic{equation}}
\setcounter{equation}{0}

The same identity (\ref{wardNgluon}) could be derived analogously for the spinor and gluon loop cases 
at the path-integral level using appropriate supersymmetric generalizations of 
(\ref{wi-N}), (\ref{defV}), (\ref{Ngluonscalarordered}) (those
representations have been summarized,  e.g., in \cite{92}). 
However, we find it more convenient to show the independence of the Ward identity of spin
by the following argument. When substituting any $\varepsilon_i$ by $k_i$ in the partially integrated integrand there are two types of terms, those where the index $i$ belongs to a cycle and those 
where not. For the first type of terms the polarization vector $\varepsilon_i$ is contained in 
the field strength tensor $f_i$, so that they get annihilated by the substitution, and this is 
independent of the application of the loop replacement rules. The second type of terms are
the ones that produce the right-hand side of the Ward identity, however since in those all
the cycle factors are unaffected by the substitution they appear as factors on both sides, 
so that again the form of the identity does not get altered by the application of the loop replacement 
rules. 

\section{Conclusions}
\label{sec:conc}
\renewcommand{\theequation}{5.\arabic{equation}}
\setcounter{equation}{0}

To summarize, we have shown that the one-loop
QCD 1PI $N$-gluon amplitudes off-shell obey the Ward identity (\ref{wardNgluon}).
This identity holds unambiguously for the scalar and spinor loop cases, but 
for the spin one case if and only if the gauge fixing is done using the
BMF with quantum Feynman gauge (or equivalently using the pinch technique).
Despite of its simplicity, to the best of our knowledge it 
has previously been given in the literature only for the four-point case \cite{papavassiliou4gluon}. 
However, an analogous identity has been derived in string theory in a similar way \cite{porrati}.

As a final comment, let us remark that the fact that the BFM 
for the gluon loop leads to the same simple, ghost-free Ward identities as for the scalar and spinor
loop only if the gluon in the loop is taken in Feynman gauge also implies that a generalization 
of the existing worldline path integral representations of the nonabelian effective action \cite{strassler1,18} to other covariant gauges must by necessity run into some algebraic complications.

\end{document}